# Projective Geometry, Duality and Plücker Coordinates for Geometric Computations with Determinants on GPUs

## Vaclav Skala

*Department of Computer Science and Engineering, Faculty of Applied Sciences, University of West Bohemia,*
*Univerzitni 8, CZ 306 14 Plzen, Czech Republic*
*http://www.VaclavSkala.eu*

**Abstract.** Many algorithms used are based on geometrical computation. There are several criteria in selecting appropriate algorithm from already known. Recently, the fastest algorithms have been preferred. Nowadays, algorithms with a high stability are preferred. Also today's technology and computer architecture, like GPU etc., plays a significant role for large data processing. However, some algorithms are ill-conditioned due to numerical representation used; result of the floating point representation.

In this paper, relations between projective representation, duality and Plücker coordinates will be explored with demonstration on simple geometric examples. The presented approach is convenient especially for application on GPUs or vector-vector computational architectures.

## INTRODUCTION

There are many algorithms based on interpolation, approximation and intersection of geometrical objects computation. Also many applications use linear algebra and solution of a linear system of equations $Ax = b$ or $Ax = 0$. A solution of such problems is made usually in the Euclidean space. However, there is a possibility to formulate problems in the projective space using homogeneous coordinates. In many cases such approach leads to more stable algorithms as projective representation enables to represent also "infinity". Also some problems the principle of duality can be used to solve the problem easily or have a one algorithm for a problem and its dual problem.

Let us consider, for the simplicity, a point $X = (X, Y) \in E^2$ and its representation in the projective space using homogeneous coordinates $x = [x, y: w]^T \in P^3$, where $w \neq 0$ is called homogeneous coordinate. If $w = 0$ the point is in infinity and it should be considered as a vector. The conversion from the projective space to the Euclidean space is given as

$$X = \frac{x}{w} \qquad\qquad Y = \frac{y}{w} \qquad\qquad (1)$$

It should be noted that there is a different notation used $x = [x_0: x_1, \ldots, x_n]^T \in P^n$ and $x_0$ is the homogeneous coordinate, now. It can be seen that the homogeneous coordinate $w$ is just a multiplicative constant without a physical unit, while the coordinates $x, y$ have some physical units, e.g. [m] or [mm] etc.

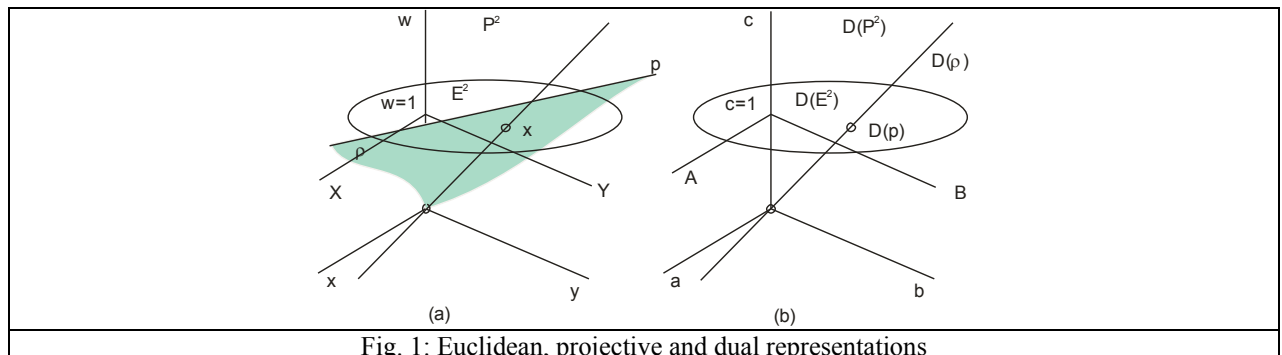

Fig. 1: Euclidean, projective and dual representations

There are two significant outcomes of the projective representation:

- instead of $n$ division operations in the $P^n$ case, only one multiplication of the homogeneous coordinate $w$ is needed,

- precision of computation, resp. representation, is nearly doubled as mantissa of the $x, y$ and $w$ are used.

It should be noted, that some exponent normalization, actually their subtraction, might be needed as exponents might tend to overflow or underflow.

A point is defined in the Euclidean space $E^2$ by coordinates $X = (X, Y)$ or as a point in the projective space $P^2$ with homogeneous coordinates $x = [x, y:w]^T$, where $w = 1$ usually. The point $x$ is actually a "line" without the origin in the projective space $P^2$, and $X = x/w$ and $Y = y/w$. It can be seen that a line $p \in E^2$ is actually a plane $\rho \in P^2$ without the origin in the projective space, i.e. the Euclidean line $p$ is defined as:

$$ax + by + cw = 0 \qquad w \neq 0 \qquad (2)$$

Any $\xi \neq 0$ can multiply the equation without any effect to the geometry due to implicit formulation. In dual representation, Fig.1.b, the plane $\rho$ can be represented as a line $D(\rho) \in D(P^2)$ or as a point $D(p) \in D(E^2)$, when a projection is made, e.g. for $c = 1$. Complete theory on projective spaces can be found in [Sto01a], [Cox69a].

Projective representation and its application for computation are considered to be mysterious or too complex. However, we are using it naturally frequently in the form of fractions, e.g. $a/b$. We also know that fractions help us to express values, which cannot be expressed precisely due to limited length of a mantissa, e.g. $1/3 = 0,33 \dots \dots .333. \dots = 0.\overline{3}$.

# DUALITY

The principle of duality states that any theorem remains true when we interchange the words "point" and "line" in $P^2$, resp. "point" and "plane" in $P^3$, "lie on" and "pass through", "join" and "intersection" and so on. Once the theorem has been established, the dual theorem is obtained as described above [Joh96a]. In other words, the principle of duality in $P^2$ says that in all theorems it is possible to substitute a term "point" by a term "line" and term "line" by the term "point" and the given theorem stays valid. This helps a lot in the solution of some geometrical cases, similarly in $P^3$.

It means that intersection computation of two lines in $E^2$ is dual to computation of a line in $E^2$ given by two points in $E^2$. It is strange as usual solution in the first case leads to formulation $Ax = b$ while in the second case parameters of a line are determined as $Ax = 0$, i.e. dual problems are solved differently if the Euclidean representation is used. However if the projective representation is used, both cases are solved as $Ax = 0$. In the case of $E^2$ computation of the intersection point of three planes is dual to computation of a plane given by three points.

# EXTENDED CROSS-PRODUCT

Many problems in computer vision, computer graphics and visualization are 3-dimensional. Therefore specific numerical approaches can be applied to speed up the solution. In the following extended cross-product, also called outer product or progressive product, is introduced in the "classical" notation using " $\times$ " symbol.

Let us consider the "standard" cross-product of two vectors $a = [a_1, a_2, a_3]^T$ and $b = [b_1, b_2, b_3]^T$. Then the cross-product is defined as:

$$a \times b = \det \begin{bmatrix} i & j & k \\ a_1 & a_2 & a_3 \\ b_1 & b_2 & b_3 \end{bmatrix} \qquad (3)$$

where: $i = [1,0,0]^T, j = [0,1,0]^T, k = [0,0,1]^T$.

If a matrix form is needed, then we can write:

$$a \times b = \begin{bmatrix} 0 & -a_3 & a_2 \\ a_3 & 0 & -a_1 \\ -a_2 & a_1 & 0 \end{bmatrix} \begin{bmatrix} b_1 \\ b_2 \\ b_3 \end{bmatrix} \qquad (4)$$

In some applications the matrix form is more convenient.

Let us introduce the extended cross-product of three vectors $a = [a_1, \dots, a_n]^T$, $b = [b_1, \dots, b_n]^T$ and $c = [c_1, \dots, c_n]^T$, $n = 4$ as:

$$\boldsymbol{a} \times \boldsymbol{b} \times \boldsymbol{c} = \det \begin{bmatrix} \boldsymbol{i} & \boldsymbol{j} & \boldsymbol{k} & \boldsymbol{l} \\ a_1 & a_2 & a_3 & a_4 \\ b_1 & b_2 & b_3 & b_4 \\ c_1 & c_2 & c_3 & c_4 \end{bmatrix} \quad (5)$$

where: $\boldsymbol{i} = [1,0,0,0]^T$, $\boldsymbol{j} = [0,1,0,0]^T$, $\boldsymbol{k} = [0,0,1,0]^T$, $\boldsymbol{l} = [0,0,0,1]^T$.

It can be shown that there exists a matrix form for the extended cross-product representation:

$$\boldsymbol{a} \times \boldsymbol{b} \times \boldsymbol{c} = (-1)^{n+1} \begin{bmatrix} 0 & -\delta_{34} & \delta_{24} & -\delta_{23} \\ \delta_{34} & 0 & -\delta_{14} & \delta_{13} \\ -\delta_{24} & \delta_{14} & 0 & -\delta_{12} \\ \delta_{23} & -\delta_{13} & \delta_{12} & 0 \end{bmatrix} \begin{bmatrix} c_1 \\ c_2 \\ c_3 \\ c_4 \end{bmatrix} \quad (6)$$

where: $n = 4$. In this case and $\delta_{ij}$ are sub-determinants with columns $i, j$ of the matrix $\boldsymbol{T}$ defined as:

$$\boldsymbol{T} = \begin{bmatrix} a_1 & a_2 & a_3 & a_4 \\ b_1 & b_2 & b_3 & b_4 \end{bmatrix} \quad (7)$$

e.g. sub-determinant $\delta_{24} = \det \begin{bmatrix} a_2 & a_4 \\ b_2 & b_4 \end{bmatrix}$ etc.

The extended cross-product for 5-dimensions is defined as:

$$\boldsymbol{a} \times \boldsymbol{b} \times \boldsymbol{c} \times \boldsymbol{d} = \det \begin{bmatrix} \boldsymbol{i} & \boldsymbol{j} & \boldsymbol{k} & \boldsymbol{l} & \boldsymbol{n} \\ a_1 & a_2 & a_3 & a_4 & a_5 \\ b_1 & b_2 & b_3 & b_4 & b_5 \\ c_1 & c_2 & c_3 & c_4 & c_5 \\ d_1 & d_2 & d_3 & d_4 & d_5 \end{bmatrix} \quad (8)$$

where: $\boldsymbol{i} = [1,0,0,0,0]^T$, $\boldsymbol{j} = [0,1,0,0,0]^T$, $\boldsymbol{k} = [0,0,1,0,0]^T$, $\boldsymbol{l} = [0,0,0,1,0]^T$, $\boldsymbol{n} = [0,0,0,0,1]^T$.

It can be shown that there exists a matrix form as well:

$$\boldsymbol{a} \times \boldsymbol{b} \times \boldsymbol{c} \times \boldsymbol{d} = (-1)^{n+1} \begin{bmatrix} 0 & -\delta_{345} & \delta_{245} & -\delta_{235} & \delta_{234} \\ \delta_{345} & 0 & -\delta_{145} & \delta_{135} & -\delta_{134} \\ -\delta_{245} & \delta_{145} & 0 & -\delta_{125} & \delta_{124} \\ \delta_{235} & -\delta_{135} & \delta_{125} & 0 & -\delta_{123} \\ -\delta_{234} & \delta_{134} & -\delta_{124} & \delta_{123} & 0 \end{bmatrix} \begin{bmatrix} d_1 \\ d_2 \\ d_3 \\ d_4 \\ d_5 \end{bmatrix} \quad (9)$$

where $n = 5$. In this case and $\delta_{ijk}$ are sub-determinants with columns $i, j, k$ of the matrix $\boldsymbol{T}$ defined as:

$$\boldsymbol{T} = \begin{bmatrix} a_1 & a_2 & a_3 & a_4 & a_5 \\ b_1 & b_2 & b_3 & b_4 & b_5 \\ c_1 & c_2 & c_3 & c_4 & c_5 \end{bmatrix} \quad (10)$$

e.g. sub-determinant $\delta_{245}$ is defined as:

$$\delta_{245} = \det \begin{bmatrix} a_2 & a_4 & a_5 \\ b_2 & b_4 & b_5 \\ c_2 & c_4 & c_5 \end{bmatrix} = a_2 \det \begin{bmatrix} b_4 & b_5 \\ c_4 & c_5 \end{bmatrix} - a_4 \det \begin{bmatrix} b_2 & b_5 \\ c_2 & c_5 \end{bmatrix} + a_5 \det \begin{bmatrix} b_2 & b_4 \\ c_2 & c_4 \end{bmatrix} \quad (11)$$

In spite of the "complicated" description above, this approach leads to a faster computation in the case of lower dimensions.

## LINEAR INTERPOLATION

Linear interpolation is heavily used across many mathematically oriented methods. However, if results, i.e. values of computation are in the homogeneous coordinates, the conversion to the Euclidean coordinates is made or a simple linear interpolation is used as:

$$x(t) = \frac{x_1}{w_1} + \left(\frac{x_2}{w_2} - \frac{x_1}{w_1}\right)t = \frac{x_1}{w_1} + \frac{w_1 x_2 - w_2 x_1}{w_2 w_1} t \triangleq [w_2 x_1 + (w_1 x_2 - w_2 x_1)t : w_2 w_1]^T \qquad (12)$$

and similarly for $y, z$ coordinates. It means that instead of 6 division operations used for $x, y, z$ coordinates we need only $6 + 1$ multiplications instead, resulting to speed up and higher precision in general. This approval leads to linear interpolation with linear parameterization.

However in some algorithms we actually need to determine relation, e.g. which point is closer and which is most distant etc. In this case, linear interpolation with a monotonic parameterization can be used and in this case

$$\boldsymbol{x}(t) = \boldsymbol{x_1} + (\boldsymbol{x_2} - \boldsymbol{x_1})t = \qquad (13)$$

where: $\boldsymbol{x_k} = [x_k, y_k, z_k : w_k]^T$, $k = 1,2$, i.e. we apply the scheme for the homogeneous coordinate as well.

In many algorithms we

et us

Least Square Error (TLSE) method is requited but for a case, when points are given in $E^3$ and a line $\pi \in E^3$, fitting th

# PLÜCKER COORDINATES

e TLSE criterion, is to be found. Note that this

In the following we will demonstrate that on very simple geometrical problems like intersection of two lines, resp. three planes and computation of a line given by two points, resp. of a plane given by three points.

# JOIN AND INTERSECTION COMPUTATIONS

Solution of non-homogeneous system of equation $\boldsymbol{AX} = \boldsymbol{b}$ is used in many computational tasks.

For simplicity of explanation, let us consider a simple example of intersection computation of two lines $p_1$ a $p_2$ in $E^2$ given as:

$$p_1 : A_1 X + B_1 Y + C_1 = 0 \qquad\qquad p_2 : A_2 X + B_2 Y + C_2 = 0 \qquad (1)$$

An intersection point of two those lines is given as a solution of a linear system of equations: $\boldsymbol{Ax} = \boldsymbol{b}$:

$$\begin{bmatrix} a_1 & b_1 \\ a_2 & b_2 \end{bmatrix}\begin{bmatrix} X \\ Y \end{bmatrix} = \begin{bmatrix} -c_1 \\ -c_2 \end{bmatrix} \qquad (2)$$

Generally, for the given system of $n$ liner equations with $n$ unknowns in the form $\boldsymbol{AX} = \boldsymbol{b}$ the solution is given:

$$X_i = \frac{\det(\boldsymbol{A_i})}{\det(\boldsymbol{A})} \qquad\qquad i = 1, \dots, n \qquad (3)$$

where: $\boldsymbol{A}$ is a regular matrix $n \times n$ having non-zero determinant, the matrix $\boldsymbol{A_i}$ is the matrix $\boldsymbol{A}$ with replaced $i^{th}$ column by the vector $\boldsymbol{b}$ and $\boldsymbol{X} = [X_1, \dots, X_n]^T$ is a vector of unknown values.

In a low dimensional case using general methods for solution of linear equations, e.g. Gauss-Seidel elimination etc., is computational expensive. Also division operation is computationally expensive and decreasing precision of a solution.

Usually, a condition **if** $\det(\boldsymbol{A}) < eps$ **then** EXIT is taken for solving "close to singular cases". Of course, nobody knows, what a value of $eps$ is appropriate.

**1    Solution of $\boldsymbol{Ax} = 0$**

There is another very simple geometrical problem; determination of a line $p$ given by two points $\boldsymbol{X}_1 = (X_1, Y_1)$ and $\boldsymbol{X}_2 = (X_2, Y_2)$ in $E^2$. This seems to be a quite simple problem as we can write:

$$aX_1 + bY_1 + c = 0 \qquad\qquad aX_2 + bY_2 + c = 0 \qquad\qquad (4)$$

i.e. it leads to a solution of homogeneous systems of equations $\boldsymbol{AX} = \boldsymbol{0}$, i.e.:

$$\begin{bmatrix} X_1 & Y_1 & 1 \\ X_2 & Y_2 & 1 \end{bmatrix} \begin{bmatrix} a \\ b \\ c \end{bmatrix} = \boldsymbol{0} \qquad\qquad (5)$$

In this case, we obtain one parametric set of solutions as the Eq.(5) can be multiplied by any value $q \neq 0$ and the line is the same.

There is a problem – we know that lines and points are dual in the $E^2$ case, so the question is why the solutions are not dual. However if the projective representation is used the duality principle will be valid, as follows.

## 2    Solution $\boldsymbol{Ax} = \boldsymbol{b}$ and $\boldsymbol{Ax} = \boldsymbol{0}$

Let us consider again intersection of two lines $\boldsymbol{p}_1 = [a_1, b_1 : c_1]^T$ a $\boldsymbol{p}_2 = [a_2, b_2 : c_2]^T$ leading to a solution of non-homogeneous linear system $\boldsymbol{AX} = \boldsymbol{b}$, which is given as:

$$p_1 : a_1 X + b_1 Y + c_1 = 0 \qquad\qquad p_2 : a_2 X + b_2 Y + c_2 = 0 \qquad\qquad (6)$$

If the equations are multiplied by $w \neq 0$ we obtain:

$$
\begin{aligned}
p_1 : a_1 X + b_1 Y + c_1 &\triangleq & p_2 : a_2 X + b_2 Y + c_2 &\triangleq \\
a_1 x + b_1 y + c_1 w &= 0 & a_2 x + b_2 y + c_2 w &= 0
\end{aligned}
\qquad (7)
$$

where: $\triangleq$ means „projectively equaivalent to" as $x = wX$ and $y = wY$.

Now we can rewrite the equations to the matrix form as $\boldsymbol{Ax} = \boldsymbol{0}$:

$$\begin{bmatrix} a_1 & b_1 & -c_1 \\ a_2 & b_2 & -b_2 \end{bmatrix} \begin{bmatrix} x \\ y \\ w \end{bmatrix} = \begin{bmatrix} 0 \\ 0 \end{bmatrix} \qquad\qquad (8)$$

where $\boldsymbol{x} = [x, y : w]^T$ is the intersection point in the homogeneous coordinates.

In the case of computation of a line given by two points given in homogeneous coordinates, i.e. $\boldsymbol{x}_1 = [x_1, y_1 : w_1]^T$ and $\boldsymbol{x}_2 = [x_2, y_2 : w_2]^T$, the Eq.(4) is multiplied by $w_1 \neq 0$, resp. by $w_2 \neq 0$. Then, we get a solution in the matrix form as $\boldsymbol{Ax} = \boldsymbol{0}$, i.e.

$$\begin{bmatrix} x_1 & y_1 & w_1 \\ x_2 & y_2 & w_2 \end{bmatrix} \begin{bmatrix} a \\ b \\ c \end{bmatrix} = \boldsymbol{0} \qquad\qquad (9)$$

Now, we can see that the formulation is leading in the both cases to the same numerical problem: to a solution of a homogeneous linear system of equations.

However, a solution of homogeneous linear system of equations is not quite straightforward as there is a one parametric set of solutions and all of them are projectively equivalent. It can be seen that the solution of Eq. (8), i.e. intersection of two lines in $E^2$, is equivalent to:

$$\boldsymbol{x} = \boldsymbol{p}_1 \times \boldsymbol{p}_2 \qquad\qquad (10)$$

and due to the principle of duality we can write for a line given by two points:

$$\boldsymbol{p} = \boldsymbol{x}_1 \times \boldsymbol{x}_2 \qquad\qquad (11)$$

In the three dimensional case we can use extended cross-product [12][15][16].

A plane $\rho : aX + bY + cY + d = 0$ given by three points $\boldsymbol{x}_1 = [x_1, y_1, z_1 : w_1]^T$, $\boldsymbol{x}_2 = [x_2, y_2, z_2 : w_2]^T$ and $\boldsymbol{x}_2 = [x_3, y_3, z_3 : w_3]^T$ is determined in the projective representation as:

$$\boldsymbol{\rho} = [a, b, c : d]^T = \boldsymbol{x}_1 \times \boldsymbol{x}_2 \times \boldsymbol{x}_2 \qquad\qquad (12)$$

and the intersection point $\boldsymbol{x}$ of three planes points $\boldsymbol{\rho}_1 = [a_1, b_1, c_1 : d_1]^T$, $\boldsymbol{\rho}_2 = [a_2, b_2, c_2 : d_2]^T$ and $\boldsymbol{\rho}_3 = [a_3, b_3, c_3 : d_3]^T$ is determined in the projective representation as:

$$\boldsymbol{x} = [x, y, z : w]^T = \boldsymbol{\rho}_1 \times \boldsymbol{\rho}_2 \times \boldsymbol{\rho}_2 \qquad\qquad (13)$$

due to the duality principle.

It can be seen that there is no division operation needed, if the result can be left in the projective representation. The approach presented above has another one great advantage as it allows symbolic manipulation as we have avoided numerical solution and also precision is nearly doubled.

## 3    Barycentric coordinates computation

Barycentric coordinates are often used in many engineering applications, not only in geometry. The barycentric coordinates computation leads to a solution of a system of linear equations. However it was shown, that a solution of a linear system equations is equivalent to the extended cross product [12][14]. Therefore it is possible to compute barycentric coordinates using cross product which is convenient for application of SSE instructions or for GPU oriented computations. Let us demonstrate the proposed approach on a simple example again.

Given a triangle in $E^2$ defined by points $x_i = [x_i, y_i: 1]^T$, $i = 1, ..., 3$, the barycentric coordinates of the point $x_0 = [x_0, y_0: 1]^T$ can be computed as follows:

$$\begin{matrix} \lambda_1 x_1 + \lambda_2 x_2 + \lambda_3 x_3 = x_0 \\ \lambda_1 y_1 + \lambda_2 y_2 + \lambda_3 y_3 = y_0 \\ \lambda_1 + \lambda_2 + \lambda_3 = 1 \end{matrix} \tag{14}$$

For simplicity, we set $w_i = 1$, $i = 1, ..., 3$. It means that we have to solve a system of linear equations $Ax = b$:

$$\begin{bmatrix} x_1 & x_2 & x_3 \\ y_1 & y_2 & y_3 \\ 1 & 1 & 1 \end{bmatrix} \begin{bmatrix} \lambda_1 \\ \lambda_2 \\ \lambda_3 \end{bmatrix} = \begin{bmatrix} x_0 \\ y_0 \\ 1 \end{bmatrix} \tag{15}$$

if the points are given in the projective space with homogeneous coordinates $x_i = [x_i, y_i: w_i]^T$, $i = 1, ..., 3$ and $x_0 = [x_0, y_0: w_0]^T$. It can be easily proved, due to the multilinearity, we need to solve a linear system $Ax = b$:

$$\begin{bmatrix} x_1 & x_2 & x_3 \\ y_1 & y_2 & y_3 \\ w_1 & w_2 & w_3 \end{bmatrix} \begin{bmatrix} \lambda_1 \\ \lambda_2 \\ \lambda_3 \end{bmatrix} = \begin{bmatrix} x_0 \\ y_0 \\ w_0 \end{bmatrix} \tag{16}$$

Let us define new vectors containing a row of the matrix $A$ and vector $b$ as:

$$x = [x_1, x_2, x_3, x_0]^T \qquad y = [y_1, y_2, y_3, y_0]^T \qquad w = [w_1, w_2, w_3, w_0]^T \tag{17}$$

The projective barycentric coordinates $\xi = [\xi_1, \xi_2, \xi_3: \xi_w]^T$ are given as:

$$\lambda_1 = -\frac{\xi_1}{\xi_w} \quad \lambda_2 = -\frac{\xi_2}{\xi_w} \quad \lambda_3 = -\frac{\xi_3}{\xi_w} \tag{18}$$

i.e.

$$\lambda_i = -\frac{\xi_i}{\xi_w} \qquad\qquad i = 1, ..., 3 \tag{19}$$

Using the extended cross product, the projective barycentric coordinates are given as:

$$\xi = x \times y \times w = \det \begin{bmatrix} i & j & k & l \\ x_1 & x_2 & x_3 & x_0 \\ y_1 & y_2 & y_3 & y_0 \\ w_1 & w_2 & w_3 & w_4 \end{bmatrix} = [\xi_1, \xi_2, \xi_3: \xi_w]^T \tag{20}$$

where $i = [1,0,0,0]^T$, $j = [0,1,0,0]^T$, $k = [0,0,1,0]^T$, $l = [0,0,0,1]^T$.

Similarly in the $E^3$ case, given a tetrahedron in $E^3$ defined by points $x_i = [x_i, y_i, z_i: w_i]^T$, $i = 1, ..., 3$, and the point $x_0 = [x_0, y_0, z_0: w_0]^T$:

$$\begin{matrix} x = [x_1, x_2, x_3, x_4: x_0]^T & y = [y_1, y_2, y_3, y_4: y_0]^T \\ z = [z_1, z_2, z_3, z_4: z_0]^T & w = [w_1, w_2, w_3, w_4: w_0]^T \end{matrix} \tag{21}$$

Then projective barycentric coordinates are given as:

$$\xi = x \times y \times z \times w = [\xi_1, \xi_2, \xi_3, \xi_4: \xi_w]^T \tag{22}$$

The Euclidean barycentric coordinates are given as:

$$\lambda_1 = -\frac{\xi_1}{\xi_w} \quad \lambda_2 = -\frac{\xi_2}{\xi_w} \quad \lambda_3 = -\frac{\xi_3}{\xi_w} \quad \lambda_4 = -\frac{\xi_4}{\xi_w} \tag{23}$$

i.e.

$$\lambda_i = -\frac{\xi_i}{\xi_w} \qquad\qquad i = 1, ..., 4 \tag{24}$$

## How simple and elegant solution!

The presented computation of barycentric coordinates is simple and convenient for GPU use or SSE instructions. Even more, as we have assumed from the very beginning, there is no need to convert projective values to the Euclidean notation. As a direct consequence of that is, that we are saving a lot of computational time also increasing robustness of the computation, especially due to division operation elimination. As a result is represented as a rational fraction, the precision is nearly equivalent to double mantissa precision and exponent range.

Let us again present advantages of the projective representation on simple examples.

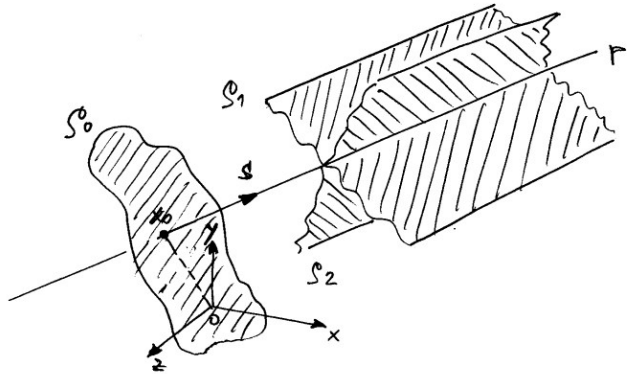

**Fig.1**: A line as the intersection of two planes

# INTERSECTION OF TWO PLANES

Intersection of two planes $\rho_1$ and $\rho_1$ in $E^3$ is seemingly a simple problem, but surprisingly computationally expensive, Fig.1. Let us consider the "standard" solution in the Euclidean space and a solution using the projective approach.

Given two planes $\rho_1$ and $\rho_2$ in $E^3$:
$$\boldsymbol{\rho}_1 = [a_1, b_1, c_1 : d_1]^T = [\boldsymbol{n}_1^T : d_1]^T \qquad\qquad \boldsymbol{\rho}_2 = [a_2, b_2, c_2 : d_2]^T = [\boldsymbol{n}_2^T : d_2]^T \qquad (25)$$
where: $\boldsymbol{n}_1$ and $\boldsymbol{n}_2$ are normal vectors of those planes.

Then the directional vector $\boldsymbol{s}$ of a parametric line $\boldsymbol{X}(t) = \boldsymbol{X}_0 + \boldsymbol{s}t$ is given by a cross product:
$$\boldsymbol{s} = \boldsymbol{n}_1 \times \boldsymbol{n}_2 \equiv [a_3, b_3, c_3]^T \qquad (26)$$
and point $\boldsymbol{X}_0 \in E^3$ of the line is given as:

$$X_0 = \frac{d_2 \begin{vmatrix} b_1 & c_1 \\ b_3 & c_3 \end{vmatrix} - d_1 \begin{vmatrix} b_2 & c_2 \\ b_3 & c_3 \end{vmatrix}}{DET} \qquad\qquad Y_0 = \frac{d_2 \begin{vmatrix} a_3 & c_3 \\ a_1 & c_1 \end{vmatrix} - d_1 \begin{vmatrix} a_3 & c_3 \\ a_2 & c_2 \end{vmatrix}}{DET}$$

$$Z_0 = \frac{d_2 \begin{vmatrix} a_1 & b_1 \\ a_3 & b_3 \end{vmatrix} - d_1 \begin{vmatrix} a_2 & b_2 \\ a_3 & b_3 \end{vmatrix}}{DET} \qquad\qquad DET = \begin{vmatrix} a_1 & b_1 & c_1 \\ a_2 & b_2 & c_2 \\ a_3 & b_3 & c_3 \end{vmatrix} \qquad (27)$$

It can be seen that the formula above is quite difficult to remember and its derivation is not simple. It should be noted that there is again a severe problem with stability and robustness if a condition like $|DET| < eps$ is used. Also the formula is not convenient for GPU or SSE applications. There is another equivalent solution based on Plücker coordinates and duality application, see [12] [16].

Let us explore a solution based on the projective representation explained above.
Given two planes $\rho_1$ and $\rho_2$. Then the directional vector $\boldsymbol{s}$ of their intersection is given as:
$$\boldsymbol{s} = \boldsymbol{n}_1 \times \boldsymbol{n}_2 \qquad (28)$$
We want to determine the point $\boldsymbol{x}_0$ of the line given as an intersection of those two planes. Let us consider a plane $\rho_0$ passing the origin of the coordinate system with the normal vector $\boldsymbol{n}_0$ equivalent to $\boldsymbol{s}$, Fig.1. This plane $\rho_0$ is represented as:
$$\boldsymbol{\rho}_0 = [a_0, b_0, c_0 : 0]^T = [\boldsymbol{s}^T : 0]^T \qquad (29)$$
Then the point $\boldsymbol{x}_0$ is simply determined as an intersection of three planes $\rho_1, \rho_2, \rho_0$ as:
$$\boldsymbol{x}_0 = \boldsymbol{\rho}_1 \times \boldsymbol{\rho}_2 \times \boldsymbol{\rho}_0 = [x_0, y_0, z_0 : w_0]^T \qquad (30)$$

It can be seen that the proposed algorithm is simple, easy to understand, elegant and convenient for SEE and GPU applications as it uses vector-vector operations.

problem formulation is different from the TLSE formulation.

et us consider a problem, when the given points $\{\langle x_p \rangle\}_{p=1}^{N}$, $x_p \in E^d$ then a line $\pi$, we are looking for, has to pass the origin of the coordinate system as:

$$x_T = \frac{1}{2N}\sum_{p=1}^{N}(x_p + x_{N+p}) = \frac{1}{2N}\sum_{p=1}^{2N}(x_p - x_p) = \mathbf{0} \tag{14}$$

as $x_p = -x_{N+p}$ ; the original data set of given points was extended. If the given points are shifted, so that $x_T = \mathbf{0}$, the given set of points $\mathbf{\Omega}$ can be used directly.

## GEOMETRIC ALGEBRA

s area $A$ of the triangle, see Fig.2, is given as:

$$2A = \left\| s \times (x_p - x_A) \right\| = \|s\|\, d_p \tag{15}$$

then the square of the distance $d_p^2$ is given as:

$$d_p^2 = \frac{\left(s \times (x_p - x_A)\right)^T \left(s \times (x_p - x_A)\right)}{s^T s} \tag{16}$$

## INTERSECTION OF TWO PLANES

As $x_T = x_A = \mathbf{0}$, we get a total o
m of Eq.**Chyba! Nenalezen zdroj odkazů.** we obtain the required line $\pi$, as the $x_A = \mathbf{0}$.

Note, that if the given points were shifted, the point $x_A$ is to be shifted back as well.

The Eq.**Chyba! Nenalezen zdroj odkazů.** is not actually dependent on a given data dimensionality, but it is dependent on a dimensionality of a geometric primitive we are mapping to. So there is an open question, e.g. how to compute TSLE in 4-dimensional case if mapping is to be made to a plane in the 4-dimensional case.

## CONCLUSION

This paper presents

a new formulation for the Total (Orthogonal) Least Square Error method for the case when points given in the $E^3$ space and a line $\pi$, fitting the TLSE criteria, is to be found. The derivation of the final formula is simple. Generalization of the presented approach to the $d$-dimensional case is straightforward. On the other hand, cases when data are given in the $n$-dimensional space and hyperplane in $k$-dimensional space is searched is solved.

In future work the given approach is to be extended to general case, when data are given in the $n$-dimensional space and hyperplane in $k$-dimensional space, where $k < n$, is searched using geometry algebra approach.

# ACKNOWLEDGMENT

The author thanks to colleagues and students at the University of West Bohemia for recommendations, constructive discussions, and hints that helped to finish the work. Many thanks belong to the anonymous reviewers for their valuable comments and suggestions that improved this paper significantly. This research was supported by the Ministry of Education of the Czech Republic – project No.LH12181.